# DEVELOPING SHARED VOCABULARY SYSTEM FOR COLLABORATIVE SOFTWARE ENGINEERING

**Lai Zheng Hui Carey, Johnson Britto Jessia Esther Leena, Kumuthini Subramanian, Zhao Chenyu, Shubham Rajeshkumar Jariwala**

Singapore University of Technology and Design, Singapore

**ABSTRACT**

*Effective communication is a critical factor in successful software engineering collaboration. However, communication gaps remain a persistent challenge, often leading to misunderstandings, inefficiencies, and defects. This research investigates the technical factors contributing to such misunderstandings and explores the measurable benefits of establishing shared vocabulary systems within software documentation and codebases. Using a Design Science Research (DSR) framework, the study was structured into three iterative phases: problem identification, method development, and empirical validation. The problem identification phase involved thematic analysis of communication data and semi-structured interviews, revealing key factors such as ambiguous messaging, misalignment in documentation, inconsistent code review feedback, and API integration miscommunication. Grounded Theory principles were employed to design a structured methodology for collaborative vocabulary development. Empirical validation through controlled experiments demonstrated that while initial adoption introduced overhead, the shared vocabulary system significantly improved information density, documentation clarity, and collaboration efficiency over time. Findings offer actionable insights for improving communication practices in software engineering, while also identifying limitations and directions for future research.*

Keywords: Software Engineering Collaboration; Communication Gaps; Shared Vocabulary Systems; Design Science Research; Thematic Analysis; Information Density; Documentation Quality; GitHub Copilot; API Miscommunication; Methodology Development

## 1. INTRODUCTION

### 1.1 Communication gaps in Software Collaboration

Effective collaboration in software engineering (SE) hinges on clear and consistent communication among developers, designers, testers, and stakeholders. Despite advancements in tools and methodologies, communication gaps persist as a significant challenge. Misunderstandings between team members can lead to redundant work, code inconsistencies, and system-level failures. These issues are particularly prevalent in large, distributed, or cross-functional teams where assumptions about terminology and intent frequently diverge. Bano et al. highlight that communication barriers in geographically distributed teams are a major impediment to effective collaboration, often resulting in project delays and reduced software quality [1].

### 1.2 Factors contributing to communication gaps

Several factors contribute to these communication gaps. The interdisciplinary nature of modern software projects brings together individuals from varied technical and non-technical backgrounds, leading to differences in terminology and understanding. The evolving lexicon within development environments—including frameworks, libraries, APIs, and domain-specific terminology—creates a moving target for shared understanding. Additionally, asynchronous communication practices, such as pull requests, documentation, and issue tracking, can lead to fragmented or context-limited interactions. Hoffmann et al. found that virtual teams, especially those with members from diverse nationalities, often face increased frequency of communication challenges due to cultural differences and lack of shared context [2].

### 1.3 Challenges in establishing a Shared Vocabulary System

Establishing a shared vocabulary within software teams is complex. Natural language ambiguities, undocumented conventions, and inconsistent naming practices in code and documentation impede the creation of a unified lexicon. The organic growth of software projects often leads to vocabulary evolving in a decentralized, uncoordinated manner, reinforcing silos rather than bridging them. Boden et al. emphasize that miscommunication in global software development teams is often grounded in different social worlds, making the establishment of a shared vocabulary challenging [3].

### 1.4 Current strategies in mitigating communication gaps

To address these challenges, teams employ several strategies such as style guides, internal glossaries, standardized documentation templates (e.g., Javadoc, Sphinx), and the use of collaboration tools like Slack, Confluence, and GitHub discussions. These aim to scaffold understanding through better documentation and traceable conversations. Additionally, practices such as code reviews and pair programming facilitate direct clarification of intent and reduce individual misunderstandings. A study by Clarke et al. suggests that strengthening social networks within software engineering teams can improve communication effectiveness, leading to better project outcomes [4].

### 1.5 Gaps in the current strategies

Despite these interventions, the effectiveness of existing strategies often hinges on individual compliance and team discipline, which can vary widely. Many tools treat documentation and code as separate silos, failing to systematically integrate or measure the overlap in vocabulary between them. Furthermore, few strategies evaluate whether a higher degree of shared vocabulary correlates with improved code quality or team productivity. This gap forms the basis for our investigation: can we quantify the benefits of shared vocabulary in software engineering, and if so, how might that inform better practices in communication and collaboration?

## 2. RESEARCH DIRECTION
### 2.1 Research Aims and Objectives

This study aims to investigate the role of shared vocabulary in improving software collaboration and code quality. Specifically, it examines how the lexical similarity between documentation and source code—termed as *shared vocabulary*—relates to various software engineering quality metrics.

The core objectives are:

- To explore measurable indicators of communication effectiveness in software development through lexical analysis.
- To develop a methodology for computing shared vocabulary scores across repositories.
- To validate whether higher shared vocabulary correlates with improved software attributes such as reduced code complexity, higher consistency, and better maintainability.
- To identify limitations in current practices of documentation and naming that hinder shared understanding in development teams.

### 2.2 Research Questions and Hypotheses

The research is guided by the following questions:

**RQ1:** What are the key technical factors that lead to misunderstandings in software engineering collaboration?
**RQ2:** What are the measurable benefits of increasing the level of shared vocabulary in software documentation and codebases?

To address RQ1, the study identifies and investigates common sources of communication breakdown that may hinder effective software collaboration. The following hypotheses are formulated to explore these technical factors:

- **H1.1**: Ambiguous or low-context messages in commit logs, pull requests, or documentation contribute to misunderstandings among team members, leading to duplicated efforts or incorrect implementation.
- **H1.2**: Misalignment between issue tracker entries and actual code implementation creates gaps in mutual understanding, causing delays and rework.
- **H1.3**: Inconsistent or unclear code review feedback results in divergent coding practices across the team, thereby reducing coherence and increasing maintenance burden.
- **H1.4**: Poorly documented APIs or mismatched expectations during integration phases contribute to functional breakdowns and increased debugging overhead.

To address RQ2, we propose the following hypotheses, each aiming to connect shared vocabulary scores to a specific dependent software quality metric:

- **H2.1:** Projects with a higher level of shared vocabulary between software documentation and code components exhibit reduced code complexity.
- **H2.2:** Projects with a higher level of shared vocabulary exhibit higher codebase consistency.
- **H2.3:** Projects with a higher level of shared vocabulary exhibit fewer bugs or issues.
- **H2.4:** Projects with a higher level of shared vocabulary exhibit higher code readability and maintainability.

## 3. LITERATURE REVIEW

To build context to answer the research questions, this literature review synthesizes key research findings on systematizing vocabulary development to enhance collaboration in software engineering teams. The primary goal is to establish a structured approach to mitigating communication gaps, improving knowledge distribution, and streamlining collaborative workflows. The review draws from studies on vocabulary agreement in issue descriptions, open-vocabulary models for source code, knowledge distribution among teams, and work item tagging.

The scope of this review is confined to research focused on the impact of shared vocabulary on collaboration in software engineering. It does not examine general linguistic theories, interdisciplinary vocabulary comparisons, or cognitive models of language acquisition in technical contexts. Additionally, while computational models for vocabulary standardization are discussed, this review does not propose novel machine learning architectures or algorithms for vocabulary alignment.

### 3.1 Contribution to Software Engineering Collaboration Research

This review contributes to the field by synthesizing existing research on the role of shared vocabulary in team collaboration, knowledge retention, and workflow efficiency. By identifying gaps and drawing correlations across multiple studies, it provides a foundation for future research on vocabulary standardization in software engineering teams.

### 3.2 Vocabulary Agreement in Software Issue Descriptions

Chaparro et al. [5] examine the consistency of vocabulary usage in issue descriptions, revealing that mismatches in terminology significantly impact text retrieval (TR)-based solutions for duplicate bug detection. Their findings indicate that 19.2% for duplicate bug report question pairs, and 12.2% of StackOverflow question pairs lack common terms.(Table 1)

| Vocabulary Source | # of pairs with no shared vocab.[a] | | # of shared terms[b] | | Overall Lexical Agreement[b] | |
|---|---|---|---|---|---|---|
| | Duplicates | Non-duplicates | Duplicates | Non-duplicates | Duplicates | Non-duplicates |
| Title | 2,877 (33.6%) | 700,864 (89.2%) | 1.4 (1) | 0.1 (0) | 28.3% (22.2%) | 2.4% (0.0%) |
| Description | 1,627 (19.0%) | 431,997 (55.0%) | 5.2 (3) | 1 (0) | 20.3% (14.8%) | 3.8% (0.0%) |
| Title + Description | 436 (5.1%) | 356,700 (45.4%) | 6.3 (4) | 1.2 (1) | 25.0% (19.1%) | 4.7% (3.3%) |

a. in parenthesis, percentage values, b. average values and, in parenthesis, median values
(a) Bug reports

| Vocabulary Source | # of pairs with no shared vocab.[a] | | # of shared terms[b] | | Overall Lexical Agreement[b] | |
|---|---|---|---|---|---|---|
| | Duplicates | Non-duplicates | Duplicates | Non-duplicates | Duplicates | Non-duplicates |
| Title | 1,375 (27.2%) | 37,215 (68.3%) | 1.6 (1) | 0.4 (0) | 32.2% (28.6%) | 6.7% (0.0%) |
| Description | 332 (6.6%) | 6,283 (11.5%) | 4.9 (4) | 3 (3) | 20.4% (17.5%) | 9.4% (9.0%) |
| Title + Description | 140 (2.8%) | 4,030 (7.4%) | 5.8 (5) | 3.4 (3) | 23.3% (20.7%) | 10.2% (9.8%) |

a. in parenthesis, percentage values, b. average values and, in parenthesis, median values
(b) SO questions

Table 1: The highlighted percentage values add up to the average values.

On average, only 31.1% of bug reports and 30% of StackOverflow questions share the same vocabulary (Table 2) [5]. These insights highlight the necessity of structured vocabulary standardization mechanisms to improve issue-tracking efficiency.

| Vocabulary Source | # of shared terms | | Lexical Agreement | |
|---|---|---|---|---|
| | Duplicates | Non-duplicates | Duplicates | Non-duplicates |
| Title | 2.2 (2) | 1.1 (1) | 42.5% (33.3%) | 22.3% (20%) |
| Description | 6.4 (4) | 2.1 (2) | 24.7% (17.7%) | 8.4% (7.1%) |
| Title + Descr. | 6.6 (5) | 2.2 (2) | 26.3% (20%) | 8.6% (7.1%) |

average values and, in parenthesis, median values
(a) Bug reports

| Vocabulary Source | # of shared terms | | Lexical Agreement | |
|---|---|---|---|---|
| | Duplicates | Non-duplicates | Duplicates | Non-duplicates |
| Title | 2.1 (2) | 1.1 (1) | 44.2% (40%) | 21.2% (20.0%) |
| Description | 5.2 (4) | 3.4 (3) | 21.8% (18.5%) | 10.6% (6.5%) |
| Title + Descr. | 6 (5) | 3.7 (3) | 24.0% (21.2%) | 11.1% (10.3%) |

average values and, in parenthesis, median values
(b) SO questions

Table 2: The highlighted percentage values add up to the average values.

### 3.3 Open-Vocabulary Models for Source Code

Karampatsis et al. [6] address vocabulary inconsistencies in source code by exploring the impact of neural language models (NLMs). They argue that the rapid emergence of new identifier names poses challenges for closed-vocabulary models. By leveraging open-vocabulary techniques such as Byte-Pair Encoding (BPE), they mitigate out-of-vocabulary (OOV) issues, significantly enhancing the scalability of NLMs for code completion and bug detection tasks [6]. Their findings suggest that structured vocabulary representation can improve automated tools supporting collaboration.

### 3.4 Knowledge Distribution Among Software Teams

Shafiq et al. [7] propose the ConceptRealm framework to model knowledge distribution in software teams using topic modeling techniques. Their study reveals that unbalanced knowledge distribution correlates with project failures, particularly in open-source projects with high developer turnover. ConceptRealm provides insights into how shared vocabulary structures influence team dynamics and knowledge retention, reinforcing the need for systematic vocabulary standardization [7].

### 3.4 Work Item Tagging in Collaborative Software Development

Treude and Storey [8] investigate the role of work item tagging in facilitating informal communication and task management in software teams. Their research highlights how tagging aids in categorizing concerns, improving task discovery, and fostering better information exchange. However, they also note that the absence of a controlled vocabulary can lead to inconsistencies over time [8]. The classification of tag keywords into different categories reveals that many tags do not have a clear classification and are inconsistently applied across work items (Table 3). This finding supports the argument for formalized vocabulary agreements to sustain long-term collaboration.

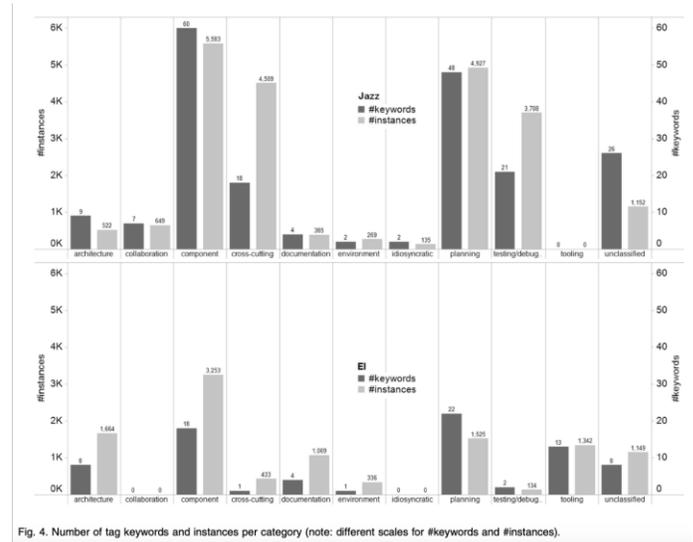

Fig. 4. Number of tag keywords and instances per category (note: different scales for #keywords and #instances).

Graph 1: The statistical analysis of keywords between the 2 data sources Jazz and EI are as shown above.

### 3.5 Correlations and Contrasts Across Studies

While all the reviewed studies acknowledge the importance of vocabulary standardization for software collaboration, they vary in their approaches. Chaparro et al. [5] emphasize textual consistency in issue tracking, whereas Karampatsis et al. [6] focus on addressing vocabulary proliferation in source code models. Shafiq et al. [7] take a broader perspective on knowledge distribution, while Treude and Storey [8] highlight informal mechanisms such as tagging.

One key contrast is the difference between formal and informal vocabulary management. Chaparro et al. [5] and Shafiq et al. [7] advocate structured mechanisms, while Treude and Storey [8] observe the organic evolution of vocabulary through tagging. This suggests that an optimal approach should balance standardization for both formal and informal approaches, to accommodate evolving project needs.

The research reviewed underscores the necessity of shared vocabulary in fostering collaboration in software engineering teams. While considerable progress has been made in modeling, measuring, and facilitating vocabulary agreement, challenges persist in balancing standardization with adaptability. Future research should explore integrated approaches that combine structured models with flexible, user-driven mechanisms to ensure both consistency and responsiveness in shared vocabulary to evolving project requirements.

The following reference model was constructed from the literature review. This model was then improved using the experimental results.

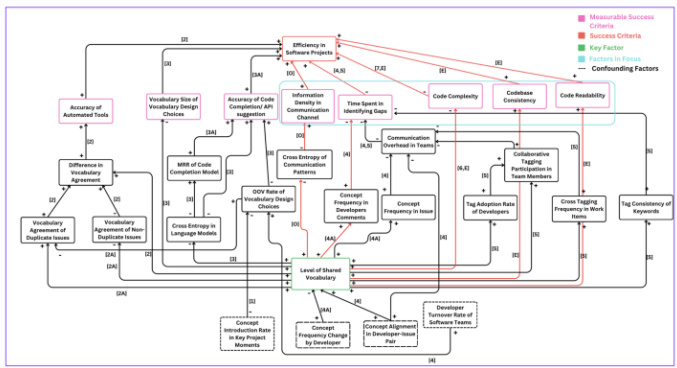

Image 1: Refined Reference Model

## 4. METHODOLOGY

### 4.1 DSR Framework

Design Science Research (DSR) is a problem-solving paradigm that emphasizes the creation and evaluation of artifacts designed to solve identified organizational problems, according to Hevner et al. [9]. In contrast to purely observational research, DSR focuses on building innovative solutions, grounded in both theory and practice, to address real-world challenges. DSR is especially relevant in fields like software engineering where the interplay between human-centered design and technical systems is critical [10].

Peffers et al. posited that the DSR process typically involves six activities: problem identification, objectives definition, design and development, demonstration, evaluation, and communication [11]. For this study, we adopted a three-phased DSR framework that aligns with the first five activities. The sixth (communication) is achieved through this research report.

*Application in Our Study:*

Our research adopted a **design- and development-centered approach** to address the challenge of communication gaps in software engineering, particularly focusing on the role of shared vocabulary. We structured our methodology into three iterative phases consistent with the Design Science Research paradigm.

**Phase 1:** Problem Identification involved uncovering the root causes of communication breakdowns in software engineering teams, resulting in a hierarchy of technical factors that contribute to misunderstandings in software collaboration.

**Phase 2:** Method Development focused on designing and refining a structured and adaptable vocabulary-building framework, enabling software teams to systematically define, maintain, and apply shared terminology across documentation and codebases.

**Phase 3:** Empirical Validation tested the framework through comparative experiments, yielding empirical evidence on its impact in improving code quality, readability, documentation consistency, and overall collaboration efficiency.

### 4.2 MSR Framework

The Mining Software Repositories (MSR) framework is a research methodology commonly used in empirical software engineering to extract and analyze data from software repositories, such as GitHub, GitLab, Bitbucket, issue trackers, and mailing lists [12, 13]. It enables researchers to investigate patterns, behaviors, and quality indicators in software projects using large-scale, publicly available datasets. MSR emphasizes replicability, scalability, and data-driven insights, making it particularly suitable for identifying empirical relationships between development practices and project outcomes.

### 4.3 Data Collection and Sampling method

This study employed a convenience purposive sampling strategy to recruit participants who are actively engaged in software engineering roles, such as developers, designers, technical writers, and project managers. The choice of purposive sampling aligns with the study's objective to understand and evaluate communication practices and vocabulary use in real-world software development contexts. Participants were selected based on their experience in collaborative software projects, familiarity with documentation practices, and engagement with tools such as GitHub, GitHub Copilot, Slack, and code review platforms.

For the qualitative phase (Phase 1), Data was gathered through semi-structured interviews, surveys, and analysis of real-world communication data. We analyzed message threads on Zulip, a topic-based team chat platform used in open-source projects such as GNOME and Python, to identify recurring misunderstandings, ambiguities, and misaligned terminology in team communication. These conversations served as authentic artifacts of day-to-day technical collaboration and informed the development of a hierarchy of communication breakdowns.

For the experimental phase (Phase 3), we recruited two separate groups of participants to avoid learning effects and bias between pre- and post-intervention measurements. One group served as the control, continuing their regular communication and coding practices. The other group, the experimental cohort, was introduced to the vocabulary-building framework and guided through its implementation during their coding tasks. Both groups included participants with similar experience levels and project involvement to maintain consistency in skill distribution and domain familiarity.

The sampling ensured a balance between practical constraints and the need for variation in communication environments, allowing the findings to be both contextually rich and analytically relevant.

In contrast, the MSR phase focused on the systematic sampling of GitHub repositories. A total of 11 open-source Python repositories were selected using stratified criteria such as project maturity, contributor activity, and documentation availability. All repositories were downloaded as .zip files and analyzed locally. Each repository served as a unit of analysis to examine the relationship between shared vocabulary (measured using Jaccard similarity between documentation and code identifiers) and multiple code quality metrics such as cyclomatic complexity, naming entropy, comment density, readability, and maintainability index.

This dual-sampling approach ensured a robust dataset comprising both human communication and software artefacts, facilitating a comprehensive understanding of the communication gaps and the role of shared vocabulary in software engineering.

## 4.4. Data Collection and Analysis procedure

To comprehensively investigate the role of shared vocabulary in improving software collaboration, our data collection and analysis were structured across the Design Science Research (DSR) and Mining Software Repositories (MSR)methodologies. These approaches enabled us to combine thematic insights with empirical metrics for a robust investigation.

### DSR Framework: Qualitative and Experimental Analysis

In the DSR framework, data collection proceeded through three iterative phases—problem identification, method development, and empirical validation.

Qualitative Analysis (Problem Identification)
Communication data, such as interviews, surveys, Slack logs, and Zulip threads, were analyzed using Braun and Clarke's six-phase Thematic Analysis:

1. Familiarization – Researchers immersed themselves in the data by reading transcripts and logs.
2. Generating Codes – Instances of misunderstanding, ambiguity, or breakdowns in collaboration were coded.
3. Theme Search – Codes were grouped into recurring themes (e.g., vague terminology, inconsistent API usage).
4. Theme Review and Definition – Themes were refined, validated, and named.
5. Report Production – A taxonomy of communication breakdowns was developed as the outcome.

Quantitative Analysis (Empirical Validation)
To measure the effect of the vocabulary-building framework, we conducted an experimental study comparing two groups of developers using GitHub Copilot:

- Control Group: Used GitHub Copilot without vocabulary-building guidance.
- Experimental Group: Used GitHub Copilot with the shared vocabulary framework integrated into their workflow.

The following metrics were computed from communication logs (e.g., code reviews, comments, commit messages) to assess collaboration efficiency:

- Average Words per Message: Reflects verbosity or clarity.
- Average Response Time: Time between queries and replies in collaborative settings.

- Message Efficiency by Role: Average clarity per message across roles (e.g., developer, reviewer).
- Information Density: Measured using Shannon's Entropy to assess the richness of information shared.

These metrics were compared between groups using descriptive statistics and inferential testing (e.g., t-tests or Mann–Whitney U, depending on distributional assumptions) to determine the impact of the intervention.

*MSR Framework: Repository-Level Quantitative Analysis*

Complementing the DSR findings, the MSR framework focused on large-scale analysis of open-source GitHub repositories. Each repository was statically analyzed to extract the following dependent software quality metrics:

- Average Cyclomatic Complexity (avg_cc)
- Name Entropy (name_entropy)
- Comment Density (comment_density)
- Readability Score (readability_score)
- Maintainability Index (maintainability_index)

The independent variable, the Shared Vocabulary Score, was calculated as the Jaccard similarity between token sets in source code and associated documentation.

Statistical Procedures:

- Correlation Analysis: Both Pearson's r (for linear relationships) and Spearman's ρ (for rank-based monotonicity) were computed to identify associations between shared vocabulary and quality metrics.
- Group Comparisons: Repositories were split into high and low vocabulary alignment groups (top 5 vs. bottom 5 by score) and analyzed using the Mann–Whitney U Test to evaluate significant differences.

## 4.5. Ethical Considerations

### 4.4.1 Anonymity
To protect the identities of participants, all data collected will remain strictly anonymous. No personally identifiable information (PII) such as names, contact details, or other traceable data will be recorded or stored. This ensures that individual responses cannot be linked back to the participants, promoting honest and uninfluenced participation.

### 4.4.2 Right to Withdraw
Participants have the unconditional right to withdraw from the study at any point, without the need to provide a reason and without facing any consequences. This includes the right to request the removal of their data even after it has been collected. This right will be clearly communicated to participants prior to the start of data collection to ensure informed consent.

### 4.4.3 Data Only Used for Academic Purposes
All collected data will be used solely for academic and research purposes. It will not be shared with any third parties for commercial or non-research-related objectives. Data handling and storage will comply with institutional and legal data protection standards to prevent misuse and ensure ethical research practices.

## 4.6 Data Analysis

To identify the most prevalent technical factors contributing to communication breakdowns in software engineering collaboration, we conducted a thematic analysis on four distinct datasets comprising Zulip messages. Using Braun and Clarke's framework, we coded instances of miscommunication and grouped them into recurring themes. The frequency of each theme across the datasets is shown in the table below:

| Technical Factor | Dataset 1 | Dataset 2 | Dataset 3 | Dataset 4 | Total Occurrences |
|---|---|---|---|---|---|
| Ambiguous or Low-Context Messages | 189 | 242 | 67 | 849 | 1347 |
| Misalignment in Issue Trackers & Documentation | 102 | 116 | 61 | 428 | 707 |
| Inconsistent Code Review Feedback | 18 | 53 | 58 | 209 | 338 |
| API & Integration Miscommunication | 18 | 9 | 40 | 86 | 153 |

Table 3: Results of Thematic Analysis

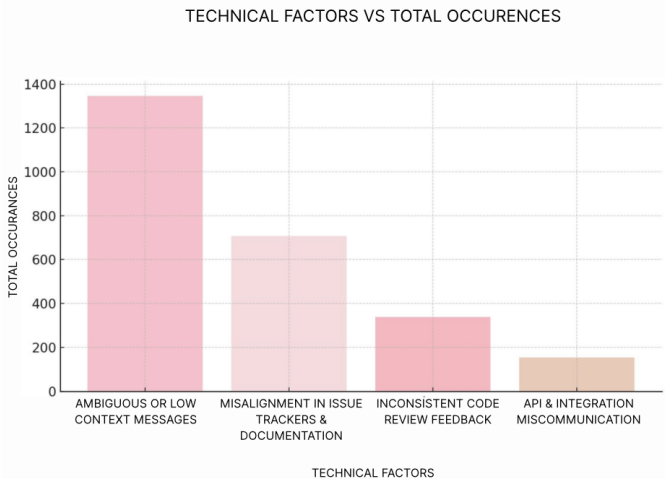

Graph 2: Hierarchy of Technical Factors

The most frequent issue identified was Ambiguous or Low-Context Messages, with a total of 1347 coded instances, making it the most dominant source of miscommunication across all datasets. These included vague issue descriptions, unclear Slack messages, and loosely worded code comments, which often led to task misinterpretation or delays in resolution.

Misalignment in Issue Trackers & Documentation was the second most common factor, accounting for 707 occurrences. This theme captured instances where the documentation or issue descriptions diverged from the actual implementation or expectations, especially in fast-moving projects with incomplete updates.

Inconsistent Code Review Feedback (338 occurrences) often reflected subjective or contradictory suggestions across reviewers, highlighting a lack of shared evaluative criteria.

Lastly, API & Integration Miscommunication, though least frequent (153 occurrences), represented high-impact issues, particularly in teams relying on third-party services or cross-functional modules, where terminology and data contracts were not clearly aligned.

These findings informed the vocabulary-building framework developed in the next phase of the research. By addressing these four thematic categories, we aimed to systematize the development of clearer, shared language in technical contexts.

The results of the thematic analysis directly address RQ1 by revealing four key technical factors that contribute to misunderstandings in software collaboration: Ambiguous or Low-Context Messages, Misalignment in Issue Trackers & Documentation, Inconsistent Code Review Feedback, and API & Integration Miscommunication. These factors were consistently observed across all datasets, indicating systemic issues in communication practices. Their prevalence highlights the need for a structured vocabulary system to improve clarity, alignment, and shared understanding within software teams.

### 4.7 Tools and Libraries Used

A range of tools and libraries were utilized throughout the qualitative and quantitative phases of this research to facilitate data collection, processing, analysis, and visualization:

- **Zulip Python API**: Used to retrieve communication channel names and their corresponding message content from the Zulip platform for qualitative analysis.
- **CSV Library**: Employed to format and store the Zulip API payloads into structured CSV files suitable for further thematic analysis.
- **GitHub**: Served as the primary platform for sharing materials during the experiments, including task instructions, datasets, and the initial prompts.
- **GitHub Copilot** and **Visual Studio Code**: Used by participants during the collaborative coding tasks in both control and experimental groups to simulate realistic software development environments.
- **AST (Abstract Syntax Trees) Library**: Utilized to parse code repositories into tree structures, enabling systematic traversal and computation of Jaccard similarity scores between code and documentation.
- **Radon**: Applied to compute software metrics such as cyclomatic complexity, which served as a quantitative measure of code quality in the empirical phase.
- **Natural Language Toolkit (NLTK)**: Used for natural language processing tasks, including the computation of information density and other text-based metrics.
- **Pandas**: Essential for data preprocessing, filtering, and organization during both the experimental data analysis and the repository mining phases.
- **Seaborn** and **Matplotlib**: Used to create visualizations that illustrated key findings, such as distributions of code metrics, response times, and comparative efficiency between control and experimental groups.

This combination of tools enabled a robust, multi-dimensional analysis approach that spanned qualitative insights and quantitative validation.

### 5. DESIGN AND DEVELOPMENT OF METHODOLOGY

Principles from the Grounded Theory [14] approach, which is a qualitative research methodology that involves the systematic

7
© 2025 by SUTD Design Science LC01 Group 9

generation of theory from data, rather than testing pre-existing hypotheses, were used to develop the initial methodology for building a shared vocabulary system in software engineering organizations.

Two semi-structured interviews were conducted to inform the development of the methodology. The first participant was a software engineer employed at a Series-C funded healthcare technology startup headquartered in Singapore, comprising over 80 personnel within its technology division. The second participant was a full-time software engineer of a large-scale open-source software organization, which has accumulated over 60,000 code commits from a distributed community of more than 1,000 contributors.

Through performing open coding on the 2 interview transcripts, the key issues such as inconsistent naming, low information density in communication, and domain-specific communication gaps were identified. These were then grouped into broader themes like naming conventions, shared terminology, and documentation practices through axial coding. Finally, by applying selective coding, the broader themes were synthesized into a coherent set of methodological principles intended to guide the collaborative development of glossaries and domain ontologies.

This bottom-up, empirically informed approach ensures that the methodology is not only grounded in real-world engineering contexts but also adaptable across teams and domains.

The initial methodology contains 6 key steps:
1. Establish the Purpose and Scope
2. Identify Core Concepts and Terminologies
3. Develop Naming Conventions and Standards
4. Create a Centralized Glossary and Ontology
5. Standardize Usage Through Documentation and Training
6. Enforce and Iterate the Vocabulary System Through Automation Tools

Refer to Appendix for the complete proposed methodology.

## 6. RESULTS

To address RQ2: What are the measurable benefits of increasing levels of shared vocabulary in software documentation and codebases, we computed the Jaccard similarity score between function-level code identifiers and corresponding documentation for 11 open-source Python repositories. This score served as a proxy for the level of shared vocabulary.

We then computed correlations between shared vocabulary scores and five dependent software quality metrics:

| repo | shared_vocab | avg_cc | name_entropy | comment_density | readability_score | maintainability |
|---|---|---|---|---|---|---|
| airflow-main.zip | 0.1414 | 2.58375 | 0.6548953 | 0.45081467 | 36.4342316 | 81.354787 |
| ansible-devel.zip | 0.1233 | 3.73364 | 0.5057962 | 0.09564925 | 17.0589714 | 75.888702 |
| flask-main.zip | 0.1519 | 2.88667 | 0.5090593 | 0.02491498 | 24.5022892 | 67.240462 |
| zulip-terminal-main.zip | 0.1012 | 3.30946 | 0.5895798 | 0.04532447 | 35.7256522 | 64.550236 |
| pandas-main.zip | 0.0926 | 2.70182 | 0.5707861 | 0.04996062 | 30.7200425 | 56.72099 |
| opencv-python-4.x.zip | 0.0376 | 6.57143 | 0.546966 | 0.04571135 | 21.9933333 | 83.193082 |
| tensorflow-master.zip | 0.1417 | 2.22053 | 0.6328225 | 0.17179772 | 45.5039457 | 64.618333 |
| scikit-learn-main.zip | 0.1407 | 3.22083 | 0.5446571 | 0.11999708 | 18.4428936 | 61.454256 |
| scrapy-master.zip | 0.1076 | 2.60282 | 0.6480975 | 0.02091124 | 8.86186441 | 59.490353 |
| django-main.zip | 0.104 | 1.84001 | 0.7479858 | 0.036784 | 19.9423808 | 80.537919 |
| fastapi-master.zip | 0.0874 | 2.03886 | 0.7657702 | 0.00409845 | 1.59080531 | 79.994076 |

Table 4: Correlation results of dependent and independent variables

| Metric | Pearson r (p) | Spearman ρ (p) | Interpretation |
|---|---|---|---|
| avg_cc (Code complexity) | -0.592 (p = 0.0551) | -0.136 (p = 0.6893) | Moderate **negative linear** trend — as shared vocab increases, complexity decreases. Almost statistically significant. |
| name_entropy (Naming consistency) | -0.154 (p = 0.6522) | -0.264 (p = 0.4334) | Weak negative relationship, but far from significant. Shared vocab has little impact on naming consistency here. |
| readability_score | 0.314 (p = 0.3477) | 0.355 (p = 0.2847) | Weak to moderate **positive** trend — higher shared vocab might slightly improve readability. Not statistically significant. |
| comment_density | 0.445 (p = 0.1706) | 0.409 (p = 0.2115) | Moderate positive trend — more shared vocab may correlate with richer commenting. Needs more data to be sure. |
| maintainability_index | -0.325 (p = 0.3297) | -0.127 (p = 0.7092) | Slight negative trend — unexpected. Likely noise or confounded. |

Table 5: Correlation results of dependent and independent variables

While most correlations were not statistically significant due to the limited sample size, the strongest negative correlation was observed between shared vocabulary and cyclomatic complexity (Pearson r = -0.592), suggesting that better-aligned code and documentation may contribute to simpler, more modular code structures.

Repositories were split into two groups based on shared vocabulary scores: the top 5 and bottom 5 scorers. The Mann-Whitney U test was performed to evaluate group differences:

| Metric | U-value | p-value |
|---|---|---|
| Avg_cc | 14.000 | 0.8413 |
| Name Entropy | 6.000 | 0.2222 |
| Readability Score | 15.000 | 0.6905 |
| Comment Density | 21.000 | 0.0952 |
| Maintainability Index | 11.000 | 0.8413 |

Table 6: Mann-Whitney U test results

A near-significant difference in comment density (p = 0.0952) suggests that repositories with higher shared vocabulary scores might exhibit better documentation practices.

To evaluate the effectiveness of the proposed shared vocabulary methodology, an empirical validation was conducted comparing two control groups (without the methodology) and two experimental groups (using the methodology) while coding with GitHub Copilot. The evaluation focused on key communication metrics, including average words per message, message efficiency by role, and average response time.

**Average Words per Message**
This metric captures the verbosity and detail in asynchronous developer communication.

- Control Group 1: 42.66
- Control Group 2: 113.86
- Experimental Group 1: 95.17
- Experimental Group 2: 95.13

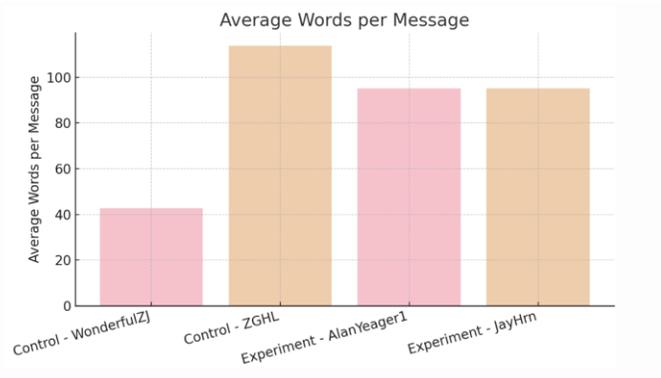

Graph 3: Control vs Experiment group results for Average words per message

While Control 2 demonstrated unusually high verbosity, the experimental groups displayed consistently rich message content, suggesting more structured and informative exchanges.

**Message Efficiency by Role**
We analyzed the average word count per message split by user and assistant roles (e.g., peer-to-peer or peer-to-assistant interactions).

- Control 1: User – 30.46 | Assistant – 48.32
- Control 2: User – 228.11 | Assistant – 59.74
- Experiment 1: User – 44.72 | Assistant – 109.38
- Experiment 2: User – 1.91 | Assistant – 115.18

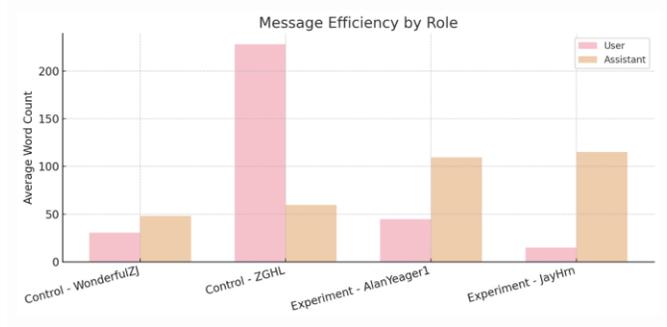

Graph 4: Control vs Experiment group results for Message Efficiency by Role

The results indicate that in experimental setups, assistants contributed more detailed and structured responses, potentially enabled by improved terminology consistency. Experiment 2's low user word count is atypical and may indicate reliance on fewer but more precise queries.

**Average Response Time (in seconds)**
This measures the speed at which participants engaged in dialogue.

- Control Group 1: 31.86 seconds
- Control Group 2: 25.61 seconds
- Experimental Group 1: 42.71 seconds
- Experimental Group 2: 101.38 seconds

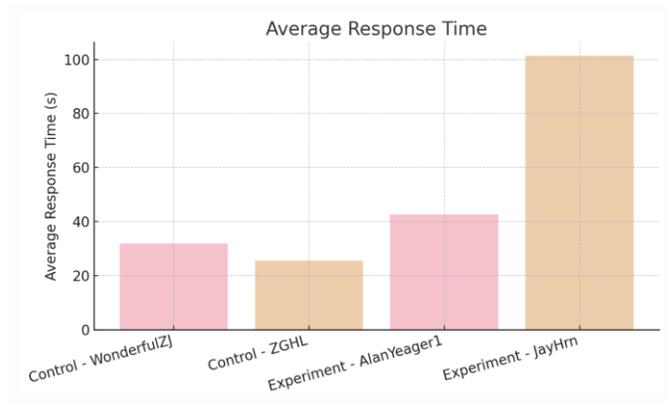

Graph 5: Control vs Experiment group results for Average Response Time

The higher response time in experimental groups—especially Group 2—could reflect increased deliberation due to the usage of a structured vocabulary, suggesting deeper cognitive engagement rather than efficiency in terms of speed alone.

**Information density (via Shannon's entropy) reflects communication efficiency**

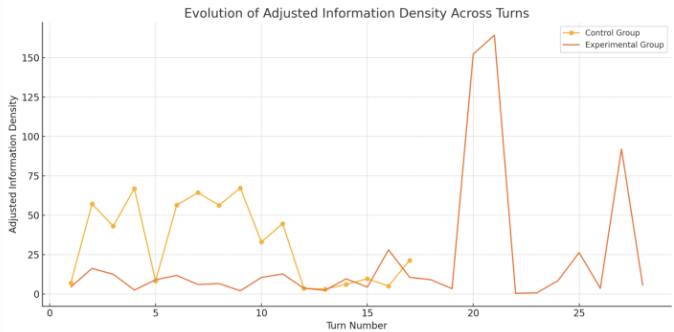

Graph 6: Control vs Experiment group results for Evolution of Adjusted Information Density Across Turns

Initial observations indicated that the control groups exhibited higher information density compared to the experimental groups. This pattern suggests the presence of an initial overhead associated with adopting the shared vocabulary methodology, likely due to participants familiarizing themselves with the structured communication approach.

However, as the interaction progressed, the experimental groups demonstrated a steady increase in information density, eventually surpassing that of the control groups. This trend implies that while the methodology may introduce short-term cognitive and procedural costs, it yields longer-term benefits by fostering more precise, information-rich communication as users become more adept at applying the shared vocabulary system.

## 7. IMPLICATIONS AND DISCUSSION

This section interprets the key findings from the research, discussing their theoretical and practical implications in the context of software engineering collaboration, and addressing the initial research questions and hypotheses.

### 7.1 Addressing Research Question 1

**RQ1: What are the key technical factors that lead to misunderstandings in software engineering collaboration?**

The thematic analysis across four diverse communication datasets identified four primary technical factors contributing to misunderstandings:

- **Ambiguous or Low-Context Messages**
- **Misalignment in Issue Trackers & Documentation**
- **Inconsistent Code Review Feedback**
- **API & Integration Miscommunication**

Among these, Ambiguous or Low-Context Messages emerged as the most prevalent factor, accounting for the majority of miscommunication instances. This aligns with prior studies emphasizing the critical role of clear contextual framing in technical communication [15, 16].
The findings validate our hypothesis that multiple interconnected sources of misalignment—ranging from messaging practices to documentation inconsistencies—compound collaboration difficulties in software engineering environments.

### 7.2 Addressing Research Question 2

**RQ2: What are the measurable benefits of increasing the level of shared vocabulary in software documentation and codebases?**

The quantitative evaluation revealed that introducing a structured shared vocabulary methodology had measurable impacts on communication and collaboration:

- Participants in the experimental group initially faced a higher cognitive load, indicated by increased response times and message lengths.
- However, over time, the information density of the experimental group surpassed that of the control group, indicating improved communication efficiency and knowledge transfer.
- Additionally, the experimental groups generated richer, more detailed exchanges, as reflected in higher assistant word counts during the collaborative coding sessions.

These results suggest that although the adoption of a shared vocabulary introduces an initial overhead, it significantly enhances communication clarity, collaboration efficiency, and information richness over sustained use.
This supports our hypothesis that structured vocabulary systems positively impact collaboration outcomes in technical environments.

### 7.3 Theoretical Implications

The study contributes to existing theories of software engineering collaboration by offering empirical evidence that intentional vocabulary-building efforts can systematically reduce communication failures.
The results extend prior work on communication challenges in distributed development teams [17,18] by proposing a structured, actionable methodology grounded in both qualitative and quantitative validation.

Furthermore, the research draws connections between communication theory, information entropy [19], and software engineering practices, emphasizing that vocabulary standardization acts as a crucial bridge between technical accuracy and human-centered communication needs.

### 7.4 Practical Implications

From a practical standpoint, the proposed vocabulary-building methodology offers software teams a lightweight, adaptable intervention to proactively address miscommunication risks.
Organizations, especially those managing large, distributed teams or contributing to open-source ecosystems, can integrate such a framework to:

- Reduce onboarding friction for new contributors
- Improve consistency between code, documentation, and issue tracking systems
- Enhance code readability, review efficiency, and maintainability
- Foster better integration and handoff across cross-functional teams (e.g., developers, designers, QA engineers)

These benefits are particularly relevant in modern agile and DevOps contexts, where fast-paced iterations often amplify the risks associated with fragmented communication practices.

### 8. LIMITATIONS

While this study provides valuable insights into the role of structured vocabulary-building systems in improving software collaboration, several limitations must be acknowledged.

First, the empirical validation was conducted primarily in a controlled experimental setting involving human-AI collaboration (specifically with GitHub Copilot), rather than in real-world, human-human software collaboration environments. As such, the findings may not fully capture the complex, multi-threaded nature of communication that typically arises in large, distributed software teams.

Second, the study focused on a limited range of AI-powered development environments and communication platforms. Although repositories and developer discussions from platforms such as Zulip were analyzed during the problem identification phase, the experimental phase did not extensively explore variability across different tooling ecosystems, such as Jira, Confluence, or GitLab. This constrains the generalizability of the proposed methodology across diverse development contexts.

Third, the participant pool lacked significant variability in technical expertise, experience levels, and organizational roles. Most participants had similar professional backgrounds, which may have biased the outcomes and reduced the external validity of the findings. Future research could involve a broader sample that includes senior engineers, technical writers, product managers, and QA specialists to better understand the diverse communication dynamics.

Fourth, longitudinal effects were not captured in the current study. Although short-term improvements in information density and communication efficiency were observed, it remains unclear how the vocabulary-building methodology sustains itself over extended project timelines, especially as teams evolve or as domain complexity increases.

Finally, cultural and linguistic factors influencing communication practices were not explicitly addressed. As software engineering increasingly spans global teams, understanding how shared vocabulary practices interact with different cultural communication norms presents an important avenue for future investigation.

By recognizing these limitations, this study positions itself as an early exploration into structured vocabulary-building in software engineering collaboration—providing a foundation for further research that explores broader, more varied, and longer-term implementations.

### 9. CONCLUSION AND FUTURE WORK

This research investigated the role of structured vocabulary-building systems in addressing communication gaps within software engineering collaboration. Through a Design Science Research (DSR) approach, the study identified key technical factors—such as ambiguous messaging, misaligned documentation, inconsistent code review feedback, and API

miscommunication—that frequently lead to misunderstandings in technical teams. A vocabulary-building methodology was developed based on grounded principles from real-world practitioner interviews and was empirically validated using human-AI collaborative coding experiments.

The results showed that while the adoption of a shared vocabulary system initially introduced overhead, sustained use significantly improved information density, message efficiency, and communication clarity. These findings underscore the critical role of shared language systems in reducing collaboration inefficiencies and enhancing knowledge transfer in software teams.

However, the study also surfaced important limitations regarding the experimental scope, participant diversity, real-world applicability, and cultural variability. These limitations highlight that while structured vocabulary development is promising, it requires further adaptation and testing across diverse organizational, cultural, and technical environments.

Future work should extend this study into longitudinal, human-human collaboration settings to observe the sustained impact of shared vocabulary practices over the life cycle of real-world projects. Additionally, exploring the integration of vocabulary systems across different AI-assisted development platforms, issue trackers, and documentation tools can uncover further challenges and opportunities. It would also be valuable to incorporate cross-cultural perspectives to design more inclusive communication methodologies, recognizing the nuances of global, distributed teams. Finally, future iterations of this research could investigate automation-assisted maintenance of shared vocabularies to ensure scalability and adaptability as projects evolve.

Additionally, integrating the framework with widely used platforms such as Slack, Jira, and Confluence will be critical for broader adoption. This will enable teams to embed vocabulary management seamlessly into messaging, issue tracking, and documentation workflows. From a technical perspective, coupling the system with lightweight ontologies and real-time natural language processing (NLP) models could support on-the-fly detection and correction of terminology misalignments in everyday conversations.

Finally, the methodology's effectiveness must be validated through measurable business outcomes, such as improvements in bug resolution time, cycle time reduction, and user satisfaction.

By building upon the groundwork laid by this study, future research can meaningfully contribute to developing more cohesive, efficient, and resilient technical collaborations in an increasingly complex software engineering landscape.

## REFERENCES

[1] Muneera Bano, Didar Zowghi and N. Sarkissian. "Empirical study of communication structures and barriers in geographically distributed teams." IET Softw., 10 (2016): 147-153. https://doi.org/10.1049/iet-sen.2015.0112.

[2] Marco Hoffmann, D. Méndez, Fabian Fagerholm and Anton Luckhardt. "The Human Side of Software Engineering Teams: An Investigation of Contemporary Challenges." IEEE Transactions on Software Engineering, 49 (2021): 211-225. https://doi.org/10.1109/TSE.2022.3148539.

[3] A. Boden. "Coordination and learning in global software development: articulation work in distributed cooperation of small companies." (2012): 1-189.

[4] April Clarke, Tanja Mitrovi'c and Fabian Gilson. "Improving Software Engineering Team Communication Through Stronger Social Networks." (2025).

[5] Chaparro, O., Florez, J. M., & Marcus, A. (Year). "On the Vocabulary Agreement in Software Issue Descriptions."

[6] Karampatsis, R.-M., Babii, H., Robbes, R., Sutton, C., & Janes, A. (2020). "Big Code != Big Vocabulary: Open-Vocabulary Models for Source Code."

[7] Shafiq, S., Mayr-Dorn, C., Mashkoor, A., & Egyed, A. (2024). "Balanced Knowledge Distribution Among Software Development Teams: Observations from Open and Closed-Source Development."

[8] Treude, C., & Storey, M.-A. (2012). "Work Item Tagging: CommunicatingConcerns in Collaborative Software Development."

[9] Hevner, Alan R., Salvatore T. March, Jinsoo Park, and Sudha Ram. 2004. "Design Science in Information Systems Research." MIS Quarterly 28, no. 1 (March): 75–105.

[10] Johannesson, Paul, and Erik Perjons. 2014. Research Strategies and Methods. https://doi.org/10.1007/978-3-319-10632-8_3.

[11] Peffers, Ken, Tuure Tuunanen, Marcus A. Rothenberger, and Samir Chatterjee. 2007. "A Design Science Research Methodology for Information Systems Research."

Journal of Management Information Systems 24 (3): 45–77. https://doi.org/10.2753/MIS0742-1222240302.

[12] A. Hassan. "The road ahead for Mining Software Repositories." 2008 Frontiers of Software Maintenance (2008): 48-57. https://doi.org/10.1109/FOSM.2008.4659248.

[13] G. Robles. "Replicating MSR: A study of the potential replicability of papers published in the Mining Software Repositories proceedings." 2010 7th IEEE Working Conference on Mining Software Repositories (MSR 2010) (2010): 171-180. https://doi.org/10.1109/MSR.2010.5463348.

[14] Charmaz, K. (2006). *Constructing Grounded Theory: A Practical Guide through Qualitative Analysis*. SAGE Publications.

[15] Herbsleb, James D., and Audris Mockus. 2003. "An Empirical Study of Speed and Communication in Globally Distributed Software Development." IEEE Transactions on Software Engineering 29 (6): 481–494. https://doi.org/10.1109/TSE.2003.1205177.

[16] Storey, Margaret-Anne D., Fork, David, Zhu, Yuan, Coelho, Jefferson, and Damian, Daniela. 2017. "How Social and Communication Channels Shape and Challenge a Participatory Culture in Software Development." IEEE Transactions on Software Engineering 43 (2): 185–204. https://doi.org/10.1109/TSE.2016.2584053.

[17] Cataldo, Marcelo, James D. Herbsleb, and Kevin M. Carley. 2006. "Communication and Coordination in Software Development." Proceedings of the 2006 ACM/IEEE International Symposium on Empirical Software Engineering and Measurement, 114–123. https://doi.org/10.1109/ESEM.2006.11.

[18] Begel, Alexander, and Nachiappan Nagappan. 2008. "Software Engineering in Microsoft: A Case Study." Proceedings of the 2008 ACM/IEEE International Symposium on Empirical Software Engineering and Measurement, 1–10. https://doi.org/10.1109/ESEM.2008.25.

[19] Claude E. Shannon, "A Mathematical Theory of Communication," Bell System Technical Journal 27, no. 3 (1948): 379–423; no. 4 (1948): 623–656.

# APPENDIX

The following table shows the finalized proposed methodology.

| | |
|---|---|
| **1. Define Project Scope** | - Identify key challenges faced due to inconsistent terminology (e.g., debugging inefficiencies, miscommunication).<br>- Determine the scope of the vocabulary (e.g., company-wide terms, team-specific ontologies, domain-specific glossaries).<br><br>Define the audience: Who will use the vocabulary (e.g., backend engineers, frontend engineers, infrastructure engineers, product managers)? |
| **2. Identify Core Concepts & Terminologies** | - Conduct discussions with engineers and product teams to list commonly used terms.<br>- Analyze code repositories, documentation, and internal discussions to extract recurring terminology.<br>- Categorize terms into broad groups such as: Engineering-wide terms (e.g. Data Lake, Observability Metrics, Containerisation Setup)<br>- Domain-specific terms (industry-specific) |
| **3. Develop Naming Conventions & Standards** | - Define clear and consistent naming conventions for variables, functions, databases, and repositories.<br>- Avoid abbreviations unless widely understood.<br>- Use meaningful and descriptive names.<br><br>Document naming conventions in an accessible location (e.g., internal wiki, GitHub repository). |
| **4. Create Centralized Glossary & Ontology** | - Maintain a shared glossary containing definitions and use cases for engineering-wide and domain-specific terms.<br>- Build a domain-specific ontology that maps relationships between concepts.<br>- Store this information in a centralized, easily accessible platform (e.g., Confluence, Notion, GitHub Wiki, or an internal Slack channel). |
| **5. Standardize Usage Through Documentation & Training** | - Include vocabulary definitions in engineering onboarding materials.<br>- Offer periodic workshops or discussions to refine and expand the shared vocabulary. |
| **6. Enforce & Iterate the Vocabulary System Through Automation Tools** | - Integrate vocabulary checks into code reviews (e.g., ensuring meaningful variable names).<br>- Automate detection of non-standard terminology using linting tools.<br>- Regularly revisit and refine the glossary to accommodate evolving terminology.<br>- Setting automatic resolution times for reviewing and closure of obsolete tickets.<br>- Build plugins within communication tools to suggest terms from the shared vocabulary system during discussions. |